\documentclass[twocolumn,showpacs,aps,prb,letterpaper,superscriptaddress]{revtex4} 
\usepackage{graphicx}
\usepackage{verbatim}
\usepackage{amsfonts,amsmath,amssymb,bm}

\begin{document}
\title{Influence of band structure effects on domain-wall resistance in diluted
ferromagnetic semiconductors}
\author{R. Oszwa{\l}dowski}
\affiliation{Institute of Physics, Nicolaus Copernicus University,
Grudzi\c{a}dzka 5, PL 87-100 Toru\'n, Poland}
\author{J. A. Majewski}
\affiliation{Institute of Theoretical Physics, Warsaw University, PL 00-681
Warszawa, Poland}
\author{T. Dietl}
\affiliation{Institute of Theoretical Physics, Warsaw University, PL 00-681
Warszawa, Poland\\ Institute of Physics, Polish Academy of Sciences and ERATO
Semiconductor Spintronics Project of Japan Science and Technology Agency,
PL 02-668 Warszawa, Poland}
\date{\today}

\begin{abstract}
Intrinsic domain-wall resistance (DWR) in (Ga,Mn)As is studied theoretically
and compared to experimental results.
The recently developed model of spin transport in diluted ferromagnetic
semiconductors [Van Dorpe {\em et al.}, Phys.~Rev.~B {\bf 72}, 205322 (2005)]
is employed.
The model combines the disorder-free Landauer-B\"uttiker formalism with the
tight-binding description of the host band structure.
The obtained results show how much the spherical $4\times 4$ $\bm{ kp}$ model
[Nguyen, Shchelushkin, and Brataas, cond-mat/0601436] overestimates DWR
in the adiabatic limit, and reveal the dependence of DWR on the magnetization
profile and crystallographic orientation of the wall.
\end{abstract}

\pacs{75.47.Jn, 72.25.Dc, 75.50.Pp}

\maketitle
Recently, current-induced domain-wall displacement\cite{Yama04Yama06} and
intrinsic domain-wall resistance\cite{Tang04,Chib06} have been observed in
ferromagnetic semiconductor \mbox{(Ga,Mn)As}. These results attract a great
deal of attention as they address the question of spin dynamics in the presence
of spatially inhomogeneous magnetization texture as well as open the doors for
novel concepts of high-density memories and logic devices.

Since in (Ga,Mn)As the domain-wall (DW) width \mbox{$\pi \lambda_W \sim 20$~nm}
is much longer than the mean free path $\ell \sim 0.5$~nm, it has been found
\cite{Chib06} that disorder, inherent to carrier-controlled ferromagnetic
semiconductors, plays an essential role in determining the magnitude of
intrinsic domain-wall resistance (DWR). It can be expected that band structure
effects, particularly the spin-orbit interaction, which accounts for
magnetoelastic\cite{Diet00,Diet01} and magnetotransport\cite{Jung06} properties
of these materials, also affect the DWR. Indeed, according to recent work of
Nguyen, Shchelushkin, and Brataas,\cite{Nguy06} in the presence of the
valence-band splitting into light and heavy hole subbands,
DWR remains non-zero even in the absence of disorder and in the adiabatic limit,
$\pi \lambda_W \gg \lambda_F$, where $\lambda_F$ is the Fermi wavelength.
This conclusion, obtained within the $4\times 4$
Luttinger model,
questions the generality of the time-honored result of
Cabrera and Falicov,\cite{Cabr74} who showed that the transition probability
of the quantum spin through a DW approaches 1
in the adiabatic limit.

In this paper we implement multi-orbital tight-binding (TB) Hamiltonian into
the
Landauer-B\"uttiker formalism for spin-dependent quantum transport in the
ballistic regime.
 The TB method is known to be more universal than $\bm{ kp}$ approaches. 
It successfully describes small quantum dots and heterojunctions between 
materials with different band structures. 
This comes at the cost of a limit to size of the studied system. 
We find, however, that this limitation is not prohibitive in the case we study
here.
Actually, a variant of this theory was already applied to describe
the spin polarization of electron current in the
(Ga,Mn)As/n-GaAs Esaki-Zener diodes\cite{VanD05,Sank06} as well as to
successfully
evaluate
the magnitude of TMR in (Ga,Mn)As/GaAs/(Ga,Mn)As trilayer structures.
\cite{Sank06}

In this work we determine the magnitude of DWR for various domain-wall widths,
shapes, and crystallographic orientations. 
In particular, we find out to what extent the $4\times 4$ spherical Luttinger
model, \cite{Nguy06} known to overestimate the effect of 
the spin-orbit interaction for the relevant values of Fermi energies,
\cite{Diet01} overvalues the magnitude of DWR in the adiabatic limit. 
Furthermore, we reveal the dependence of DWR on the 
crystallographic orientation of the DW, an effect not studied so far.
Finally, we compare results of our disorder-free theory to experimentally
determined values of DWR.

The transmission matrix employed here is determined in terms of the
ex\-ten\-ded trans\-fer-ma\-trix meth\-od\cite{Vogl94} with\-in the TB
approach, with the spin-orbit coupling included.
The (Ga,Mn)As valence-band structure for small Mn content
was shown to be similar to that of GaAs.\cite{Okab01} Thus
we use $sp^3\!s^*$ parameters proposed for GaAs.\cite{Carl95}
This model reproduces correctly
the band structure of GaAs in the relevant part of the Brillouin zone.
Moreover, it takes into account 
the Dresselhaus terms, essential for the spin-dependent transport.

The presence of the
Mn ions and, therefore, of the sp-d exchange interaction, is introduced to the
TB Hamiltonian using the virtual crystal and mean-field approximations, taking
into account the appropriate weights of Ga and As orbitals in the
wavefunctions close to the center of the Brillouin
zone.\cite{Diet01,Kacm01}
For the Mn content of 5\% the hole subband splitting at the $\Gamma$
point obtained by the present method is 150~meV and 46~meV for the heavy 
and light holes respectively,
which compares favorably with the $\bm{ kp}$ values of 150 
and 50~meV.\cite{Diet01} In what follows we use $h_0$ as a
measure of magnetization, where $3h_0$
is the heavy-holes splitting, and $h_0/2 = B_G$, where $B_G$ is the splitting
parameter of Ref.~\onlinecite{Diet01}.

To study coherent transport through the DW we define three regions along the
current direction:
left and right leads, and a central multi-layer region of the length $L_0 \gg
\lambda_W$ containing the DW, where
$L_0$, taken here to be 40~nm, can be identified as the phase coherence
length of the system,
$L_0 \approx L_{\varphi}(T)$. The Mn concentration is constant throughout
the three sections. In the central region magnetization 
$\bm{ M}$ rotates, so that its deviation $\theta$ from the initial orientation
[001], taken here as the $z$-direction, changes by small steps in 
the consecutive atomic layers, according to $\cos\theta=\tanh(x/\lambda_W)$,
where $x$ is the distance from the 
DW center in the current direction. We consider the Bloch and the N\'eel DWs,
\cite{Hube01} for which the vector $\bm{ M}(0)\times \bm{ M}(\theta)$ is
along the current direction or perpendicular to it, respectively.
In addition to $180^{\circ}$ walls we study a $90^{\circ}$ Bloch DW, for
which the magnetization in the left
lead is normal to the ${\bm M}$ direction in the right lead.
Since the virtual crystal approximation restores translational invariance in
the direction
perpendicular to the current, we model the lateral extent of the system by
Born-von Karman 
boundary conditions. We assume no strain in the entire structure.

\begin{figure}[!htbp]
\resizebox{8.cm}{!}{
\includegraphics[angle=-90]{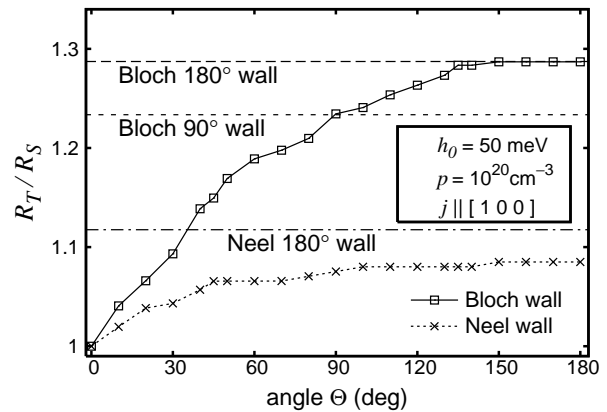}}
\caption{\label{figure_rotation}
Total resistance normalized by the Sharvin resistance, $R_{T}/R_S$, calculated
within the 
adiabatic model as a function of the angle between the magnetization and [001]
direction for the
Bloch and N\'eel walls. Horizontal lines show $R_{T}/R_S$ computed within the
full model in the
adiabatic limit ($\lambda_W/\lambda_F=1.4$) for three types of walls.
}
\end{figure}

In our calculations, we place the Fermi level $E_F$
at either 0.08~eV, 0.2~eV, or 0.36~eV below 
the mid-point of the heavy and light hole bands at the $\Gamma$ point, thus
covering the experimentally relevant 
hole concentration range, as $p=1\times 10^{20}\,\textrm{cm}^{-3}$,
$3.5\times 10^{20}\,\textrm{cm}^{-3}$, and $1\times 10^{21}\,\textrm{cm}^{-3}$,
respectively.
In order to produce a non-zero current density, we introduce a bias $V$ by
shifting the Fermi level of the
right lead with respect to the left one.
The bias value $\thicksim 10^{-4}$~eV is much smaller than typical splitting
in the valence band, so that the calculated resistances are bias-independent.
We obtain the current density from the transmission coefficients by energy
integration 
over the range from $E_F-eV$ to $E_F$ and summation over $\bm{ k}_{\parallel}$
vectors over a two-dimensional grid.\cite{Datt95} 
We find that typically grids of 200 $k$-points are sufficient, except for more
subtle problems, such as that of adiabatic passage
(Fig.~\ref{figure_rotation}), which required a 550 $k$-point grid.
Using the current density and the bias we obtain the total resistance $R_T$
of the structure containing the DW, in units $\Omega \mu\textrm{m}^2$.
We define DWR as a difference between $R_T$ and the resistance for the uniform
magnetization in the whole structure. Since disorder is neglected, the latter 
corresponds to the Sharvin 
resistance $R_S$, as all accessible channels are fully transmitting for
spatially homogeneous magnetization.

\begin{figure}[!htbp]
\resizebox{8cm}{!}{
\includegraphics[angle=-90]{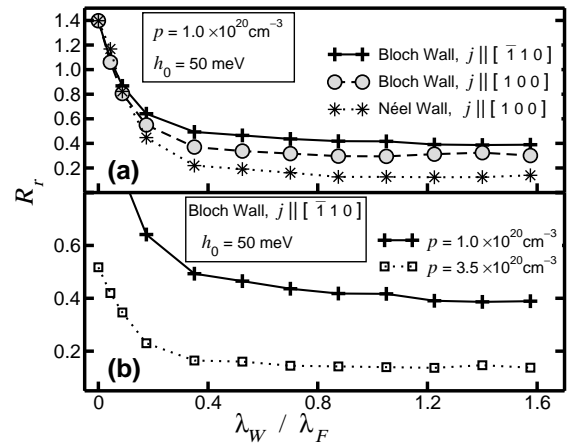}
}
\caption{\label{figure_1_and_inset}
(a) The predicted relative domain-wall resistance $R_r$ for (Ga,Mn)As
as a function of the DW width parameter $\lambda_W$
(in units of the Fermi wavelength $\lambda_F=5.7$~nm) for
various current directions and domain-wall profiles. (b) $R_r$ for 
two hole densities $p=1\times 10^{20}\,\textrm{cm}^{-3}$ 
and $p=3.5\times 10^{20}\,\textrm{cm}^{-3}$. 
}
\end{figure}

Figure \ref{figure_1_and_inset} depicts DWR obtained within our 
$sp^3\!s^*$ TB model for $h_0= 50$~meV.
We consider current flowing in either [\=110] or [100] directions, while
$\bm{ M}$ in the left (right) lead
is parallel (anti-parallel) to the [001] direction.
We present relative resistance of the DW, which is defined
as $R_r=\left(R_{T}-R_S\right)/R_S$.
In Fig.~\ref{figure_1_and_inset} the DW width $\lambda_W$ is given in units of
the Fermi wavelength $\lambda_F$ of the heavy holes for $\bm{ k}_F$ in
the [100] direction in the absence of magnetization ($h_0=0$). This corresponds
to $\lambda_F = 5.7$ and $3.7$~nm 
for $p = 10^{20}$ and $3.5\times 10^{20}$~cm$^{-3}$, respectively. 
The value $\lambda_W=0$ denotes a step-like DW profile.

We see in Fig.~\ref{figure_1_and_inset}(a) that in the adiabatic limit,
$\lambda_W\gtrsim \lambda_F$, the relative resistance
$R_r$ saturates and remains non-zero. 
The range $\lambda_W > \lambda_F$ is relevant for experiments with
typical \mbox{(Ga,Mn)As} films. 

Some residual variations of $R_r$ in this region are probably caused by
Fabry-Perot-like
interference that occurs for a finite $L_0$. At the same time, according to
the results depicted in
Fig.~\ref{figure_1_and_inset}(b), the magnitude of $R_r$ becomes significantly
smaller for higher values of the hole
density. Furthermore, $R_r$ for the N\'eel wall is seen to be substantially
lower than for the Bloch wall (Fig.~\ref{figure_1_and_inset}(a)).
These findings confirm results obtained by Nguyen, Shchelushkin, and Brataas,
\cite{Nguy06} though our more complete
theory leads to $R_r$ up to a factor of two smaller than that resulting from
the spherical $4\times 4$ model.
This can be seen for example in Fig.~\ref{figure_1_and_inset}(b), where the
lower curve corresponds to
$h_0/E_F=0.23$. In this case our TB-based model predicts $R_r\approx$0.14,
while the $\bm{kp}$ one yields almost 0.3.
Moreover, we find that the magnitude of $R_r$ varies significantly with the DW
orientation (Fig.~\ref{figure_1_and_inset}(a)). This effect results
from the dependence of $R_{T}$ on the crystallographic directions of the
current and magnetization. 

\begin{figure}[!htbp]
\resizebox{8cm}{!}{
\includegraphics[angle=-90]{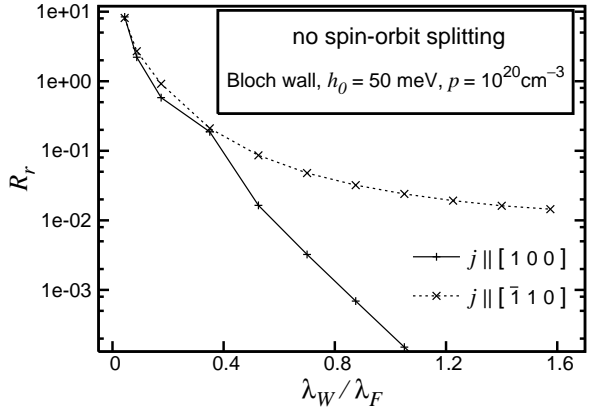}
}
\caption{\label{figure_noso2}
The theoretical predictions of intrinsic domain-wall resistance in the absence
of the
spin-orbit coupling. In the case of the domain wall normal to [1 0 0], the
relative
domain-wall resistance ($R_r$) drops exponentially to zero as a function of the
domain-wall width parameter
normalized by the Fermi wavelength. In the case of wall normal to [\=1 1 0],
the grazing incidence channels cause a slower decay of $R_r$.
Spin-orbit interaction is set to zero in this calculation,
but all other parameters remain the same as in the calculations leading to a
large value of $R_r$ in Fig.~\ref{figure_1_and_inset}. The value of $\lambda_F$
is the same as in Fig.~\ref{figure_1_and_inset}. 
}
\end{figure}

To elucidate the origin of the non-zero DWR in the adiabatic limit, we have
repeated our calculations switching off
the spin-orbit coupling. Thus, we take entire complexity of the valence band
into account except for putting 
the spin-orbit parameters $\lambda^a$ and $\lambda^c$ of the TB Hamiltonian
equal to zero in the whole structure.
As shown in Fig.~\ref{figure_noso2}, $R_r$ no longer saturates but instead
vanishes with
$\lambda_W / \lambda_F$, a behavior first noted by Cabrera and Falicov,
\cite{Cabr74} for the free electron case.
In the case of [\=110] direction the $R_r$ exhibits certain deviation from
the expected exponential drop. We ascribe this behavior to the presence of
grazing incidence channels for which the $\bm{ k}$
component along the current direction is small, so that
the reflection coefficient is still relatively high.\cite{Cabr74}
We conclude that non-zero DWR of \mbox{(Ga,Mn)As} in the adiabatic limit
originates from the spin-orbit
coupling rather than from multiband transport in the valence band. This
substantiates the previous suggestion,\cite{Nguy06}
put forward by noting that DWR vanishes if $4\times 4$ Luttinger Hamiltonian
is assumed to be diagonal.
In this case and in the thick-wall limit, the spin of the carrier can
adiabatically track the local
magnetization and the charge transport is not affected by the presence of DW.
\cite{Cabr74,Greg96Levy97}

In order to gain further understanding of the physics underlying the DW
resistance, we study the thick DW limit. 
We first consider the Bloch wall, 
for which magnetization is always perpendicular to the current, chosen here to
flow along the [100] direction. 
Owing to the symmetry of the zinc-blende structure, there is no anisotropic
magnetoresistance (AMR) in such a case.
The lack of AMR does not imply that the distribution of open channels in the
reciprocal space is identical for all magnetization directions $\theta$, 
where $\theta$ is the angle of $\bm{ M}$ with respect
 to the [001] direction. 
We show now that the dependence of this distribution on $\theta$
gives rise to DWR in the adiabatic limit. To this end we use the adiabatic
transport model, defined as follows. 
We consider all $\bm{ k}$ vectors for which carrier propagation 
for given spatially uniform $\bm{ M}\!\left(\theta\right)$ and Fermi level is
possible. 
We project these vectors on a plane
perpendicular to the current direction, obtaining a 2D closed area
$S\!\left(\theta\right)$ in the reciprocal space.
Next, we turn the magnetization by a small angle and form the projection again,
keeping constant $h_0$, $E_F$, and the tunneling direction. For each step, the
corresponding $S\!\left(\theta\right)$ is different, reflecting the rotation
of $\bm{ M}$, and the coupling between the spin and orbital degrees of freedom.
Finally, after rotating the magnetization through the half-circle,
we find a common subset of all $S\!\left(\theta\right)$ for
$0\leq\theta\leq\pi$, by retaining only these 
channels that are open for all $\theta$. 
This approach is a generalization of a qualitative model presented in
Ref.~[\onlinecite{Nguy06}].
However, we do not use the spherical approximation for the bands, so that an
$S\!\left(\theta\right)$ for a general $\theta$ cannot be obtained by a simple
rotation of $S\!\left(0\right)$.

In Fig.~\ref{figure_rotation}, we present the total DW resistance (in the
units of $R_s$) as a function of $\theta$, 
which is determined from the common subset $S_c(\theta)$ of areas $S$
corresponding
to all relevant subbands and magnetization angles up to $\theta$, and
$S\!\left(0\right)$ yields the Sharvin resistance $R_S$.
We find a very good agreement between the calculated resistance 
of the thick Bloch DW
and the one obtained in the adiabatic model.  
A direct inspection of transmission coefficients reveals that
the distribution of open channels is very similar in both approaches. 
To verify our arguments we consider a
$90^{\circ}$ DW. We find good agreement
between the calculated value of DWR and the one from the adiabatic 
model for $\theta =90^\circ$, see Fig.~\ref{figure_rotation}.

It is known that (Ga,Mn)As films reveal resistance anisotropy
when magnetization is tilted towards the current direction.\cite{Jung06,Baxt02}
The above reasoning, however, remains unchanged for the case of the N\'eel
wall, and the magnitudes of DWR can be obtained in the same way. 
In agreement with the results 
discussed above (see Fig.~\ref{figure_1_and_inset}(a)),
we find that 
more channels are left open after rotating $\bm{ M}$ by $\pi$ in the $xz$ plane
(N\'eel DW) than in the $zy$ plane (Bloch wall). 

The values of DWR calculated from the adiabatic and the
exact model for the $180^\circ$ N\'eel wall differ somewhat 
more than in the case of the Bloch DW, see Fig.~\ref{figure_rotation}.
This difference is caused by grazing incidence channels, for which 
the energy-integrated transmission coefficients are, on average, much smaller
than the ones in the case of the 
Bloch DW.  

In the limit of vanishing spin-orbit splitting, 
shapes and orientations of the areas $S$ are identical for all $\theta$ values.
 Therefore,
their common subset is equal to $S\!\left(0\right)$ and, hence, to the Sharvin
resistance $R_S$.
Thus DWR vanishes totally in the adiabatic approximation. However, when the DW
thickness decreases,
a gradual closing of the 
tunneling channels takes place. Accordingly, the magnitude of $R_r$ 
calculated within the exact model grows exponentially, 
when $\lambda_W/\lambda_F$ decreases (Fig.~\ref{figure_noso2}). 

\begin{figure}[!htbp]
\resizebox{8.cm}{!}{
\includegraphics[angle=-90]{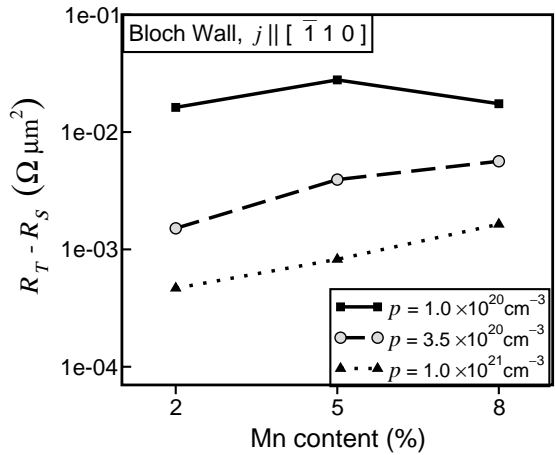}}
\caption{\label{figure_exper}
Predicted values of the  domain-wall resistance in the adiabatic
limit for various hole concentrations.
The data are for the saturation value of spin-splitting at
a given Mn concentration $x$.
}
\end{figure}
An important question arises whether the adiabatic DWR discussed here could
explain experimental magnitudes of 
DWR in \mbox{(Ga,Mn)As}. Figure~\ref{figure_exper} shows the values of
the intrinsic DWR, i.e., $R_{T} -R_S$ in the adiabatic limit for experimentally
relevant ranges of the hole concentrations and Mn densities. The computed
values of 
DWR are positive. The sign of the effect is in disagreement with the findings
of
Tang {\em et al.},\cite{Tang04} who reported a negative value of DWR for their
samples of \mbox{(Ga,Mn)As}. 
By contrast, Chiba {\em et al.}\cite{Chib06} found 
a positive value, DWR$ = 0.5\pm 0.1$~$\Omega\!\mu$m$^2$ 
for $p\approx 2\times 10^{20}$~cm$^{-3}$ and $x=0.05$.
A comparison of this experimental value with the
results of Fig.~\ref{figure_exper} shows that the computed values are at
least one order of magnitude smaller.
We recall at this point that high hole and Mn concentrations, necessary to
observe the
ferromagnetism, result in a short mean free path in these systems, typically
much smaller than the DW width.

In conclusion, according to the detailed quantitative studies presented here, 
the spin-orbit interaction leads to a non-zero value of domain-wall
resistance even 
in the adiabatic limit. The magnitude of this adiabatic resistance depends
significantly 
on the character of the domain-wall and its crystallographic orientation.
However, the sign 
and magnitude of the computed results are in variance with the recent
experimental results 
for (Ga,Mn)As. We take this disagreement as an evidence for the paramount
importance of 
disorder effects in the physics of DWR in carrier-controlled ferromagnetic
semiconductors. 
However, non-zero domain-wall resistance induced by spin-orbit interactions can
presumably be detected in those ferromagnetic systems, in which both coherence
length and mean free path are longer than the domain-wall width.

We thank P.~Kacman, H.~Ohno, and P.~Sankowski for valuable discussions.
This work was partly supported by EU NANOSPIN project (EC:FP6-2002-IST-015728).
R.O. acknowledges an additional support by Nicolaus Copernicus University Grant
381-F.

\end{document}